\RequirePackage{rotating}
\documentclass[11pt,a4paper,oneside]{article}

\usepackage[utf8]{inputenc}
\usepackage{geometry}
\usepackage{pdflscape}
\usepackage{graphicx}
\usepackage{epstopdf}
\usepackage{setspace}
\usepackage{enumerate}
\usepackage{amsmath, amsthm, amssymb}
\usepackage{mathtools}
\usepackage[longnamesfirst]{natbib}
\usepackage{subfig}
\usepackage{booktabs}
\usepackage{color}
\usepackage{soul}

\usepackage{ae} 
\usepackage[T1]{fontenc}

\DeclarePairedDelimiter{\floor}{\lfloor}{\rfloor}

\def\Prob{{\mathbb{P}}}
\def\der{{\mathrm{d}}}

\definecolor{lightblue}{rgb}{.90,.95,1}
\sethlcolor{cyan}

\title{High-Frequency Jump Analysis of the Bitcoin Market}

\author{Olivier Scaillet\footnote{University of Geneva and Swiss Finance Institute, 40 Bd du Pont d'Arve, 1211 Geneva, Switzerland. Voice: \mbox{+41 22 379 88 16}. Fax: \mbox{+41 22 379 81 04}. Email: olivier.scaillet@unige.ch. Corresponding author.} \and Adrien Treccani\footnote{University of Geneva and Swiss Finance Institute, 40 Bd du Pont d'Arve, 1211 Geneva, Switzerland. Voice: \mbox{+41 22 379 81 66}. Fax: \mbox{+41 22 379 81 04}. Email: adrien.treccani@unige.ch.} \and Christopher Trevisan\footnote{Ecole Polytechnique Fédérale de Lausanne and Swiss Finance Institute, EPFL CDM-DIR, 1015 Lausanne, Switzerland. Voice: \mbox{+41 21 693 01 28}. Email: christopher.trevisan@epfl.ch.} \footnote{Acknowledgements:  We thank participants at the "Market Microstructure and High Frequency Data" conference 2017 in Chicago for helpful comments.}}

\date{\today\\first draft: April 2017}

\begin{document}

\maketitle

\vspace*{2cm}

\begin{abstract}
We use the database leak of Mt. Gox exchange to analyze the dynamics of the price of bitcoin from June 2011 to November 2013. This gives us a rare opportunity to study an emerging retail-focused, highly speculative and unregulated market with trader identifiers at a tick transaction level. Jumps are frequent events and they cluster in time. The order flow imbalance and the preponderance of aggressive traders, as well as a widening of the bid-ask spread predict them. Jumps have short-term positive impact on market activity and illiquidity and induce a persistent change in the price. \\
\end{abstract}

JEL classification: C58, G12, G14.

Keywords: Jumps, Liquidity, High-frequency data, Bitcoin.

\newpage

{\setstretch{2}

\section{Introduction}

Bitcoin, a distributed digital currency, was created in 2009 and is the most popular cryptocurrency with a multi-billion dollar capitalization since 2013. 
It is the first such currency to gain relatively widespread adoption. The technology provides an infrastructure for maintaining a public accounting ledger and for processing transactions with no central authority.
Unlike traditional currencies, which rely on central
banks, bitcoin relies on a decentralized computer network to validate transactions and
grow money supply (see \cite{Yermack2013} and \cite{Yermack2017} for further background on the bitcoin and its technology). Each bitcoin is effectively a (divisible) unit which is transferred
between pseudonymous addresses through this network.  Its promising potential and scarcity have driven the market price of bitcoin to parity with the U.S.\ dollar in February 2011 and above \$1,000  in November 2013. It is estimated that by the end of our period of study in 2013, bitcoin had approximately one million users worldwide with a three-digit annual growth. Mt.~Gox was the largest exchange platform to provide bitcoin trading for U.S.\ dollar until it went bankrupt early 2014 as a result of the theft of client funds by hackers.\footnote{The Japanese courts are holding pre-trial hearings, and the claim process enters its fourth year. Japanese police have found part of the missing bitcoins, and the 24,000 or so claimants are waiting for a final settlement.} An important part of Mt.~Gox internal database leaked, revealing a full history of trades on the period April 2011--November 2013. This data set gives us a rare opportunity to observe the emergence of a retail-focused, highly-speculative and unregulated market at a tick frequency with trader identifiers at the transaction level.

Bitcoin has experienced numerous episodes of extreme volatility and apparent discontinuities in the price process. On one hand, the absence of solid history and exhaustive legal framework make bitcoin a very speculative investment. Because it does not rely on the stabilizing policy of a central bank, the reaction to new information, whether fundamental or speculative, results in high volatility relative to established currencies. On the other hand, the relative illiquidity of the market with no official market makers makes it fundamentally fragile to large trading volumes and to market imperfections, and thus more prone to large swings than other traded assets. The focus of this paper is to contribute to the growing literature on the analysis of jumps and their potential explanation. Jumps are sporadic events of a larger amplitude than what a continuous diffusion process can explain. Analyzing their distributional properties is important because of the consequences in applications including derivatives pricing and risk management. \citet{ChristensenOomenPodolskij14} consider U.S.\ large-cap stocks, equity indexes, and currency pairs. They conclude that jumps in financial asset prices are often erroneously identified and are, in fact, rare events accounting for a very small proportion of the total price variation. They show that measures of jump variation based on low-frequency data tend to spuriously assign a burst of volatility to the jump component. \citet{BST15} test for the presence of jumps in Dow Jones stocks at high frequency. They explain that the repetition of the jump test over a large number of days leads to a number of spurious detections because of multiple testing issues. They correct for this bias, reducing even further the number of remaining detections in comparison to the findings of \citet{ChristensenOomenPodolskij14}. They find an average of 3 to 4 remaining jumps a year and relate them to macroeconomic news, prescheduled company-specific announcements, and stories from news agencies which include a variety of unscheduled and uncategorized events. They conclude that the vast majority of news do not cause jumps but may generate a market reaction in the form of bursts of volatility. They conjecture that jumps might be related to liquidity issues and order flow imbalances but the limited number of detected jumps in their study poses a challenge for getting statistically significant empirical evidence.  

Our main contributions are to assess the presence of jumps in a highly-speculative emerging market with low liquidity, and to determine whether liquidity is a main driver of jump occurence. The information of the trader identifier and the direction of trade, i.e., whether the transaction are initiated by a buyer or a seller, provided by our data records is key for our empirical analysis. Such information is rarely available for other markets and is related to the unique way that the Mt.~Gox database stores the knowledge about successive trades. 

Our first contribution is to detect the presence of jumps in the bitcoin market, and to study their dynamics. We apply the jump detection test of \citet{LeeMykland12} to the tick data and control for multiple testing across days using the False Discovery Rate (henceforth, FDR) technique \citep[see, e.g.,][]{BST15,RomanoShaikhWolf08}. We identify 124 days including at least one jump during the period, or approximately one detection day per week. The number of detections is significantly larger than what previous research observes for large-cap assets and indices, suggesting that the intensity of jump occurrence largely varies depending on the market characteristics, such as its liquidity or the specificities of the participants. We investigate the dynamics of durations between jumps. \citet{BST15} cannot reject the hypothesis that jump arrivals follow a Poisson process. We apply a runs test on jump detections date and strongly reject the independence of inter-jump durations. Hence, jump dynamics do not support the jump process used by \citet{Merton76} and subsequent models based on compound Poisson processes with constant intensity.

Our second contribution is to perform a systematic event study for the identified jumps to characterize the market conditions preceding and following a discontinuity. We seek to determine, if not the cause, the main factors driving the occurrence of jumps as well as their impact on market conditions. Such an empirical analysis is made possible because of a sufficiently  large number of detected jumps, which is not the case for large-cap markets. We use a probit regression model and find that discontinuities are anticipated by abnormal trading activity and liquidity conditions: the order flow imbalance, the proportion of aggressive traders and the bid ask spread have significant predicting power over jumps. Those findings support the hypothesis that jumps occur when trading activity clashes with a liquidity shock, and there is no stabilizing mechanism either induced by a central bank or by market makers whose mandate is to provide liquidity. We perform a post-jump analysis of the market conditions and find that most indicators are exacerbated, including the trading volume, the number of traders, the order flow imbalance, the bid-ask spread, the realized variance, the microstructure noise variance and the proportion of aggressive traders. These factors however revert to their anterior level in less than half an hour. Comparing the price levels before and after jumps reveals a significant, persistent impact: positive (negative) jumps occur during locally bearish (bullish) trends.

The rest of the paper is organized as follows. Section 2 reviews the data and our cleaning procedure. Section 3 defines our methodology for detecting jumps. Section 4 presents our empirical results. Section 5 concludes.

\section{Data on the bitcoin market}

Let us first briefly introduce the bitcoin. Bitcoin is a novel form of electronic money that is based on a decentralised
network of participating computers. It has no physical counterpart; it is merely
arbitrary (divisible) units that exist on this network. There is no central bank and there are no interest rates. The
system has a pre-programmed money supply that grows at a decreasing rate until
reaching a fixed limit. This semi-fixed supply exacerbates volatility and deflationary
pressure.  Each user of bitcoin can generate
an address (like an email address or account number) through which to make and receive transactions,
making bitcoin pseudonymous.
The crucial aspect that makes bitcoin work is that it solves the double-spending
problem without relying on a central authority. In other words, it is possible to
send a bitcoin securely, without then being able to spend that bitcoin again,
without someone else being able to forge a transaction, and also without your
being able to claim that bitcoin back (i.e., a chargeback).
These transactions get recorded in a decentralised ledger (known as the
blockchain), which is maintained by a network of computers (called 'miners'). 
Miners maintain consensus in the blockchain through solving difficult
mathematical problems, and are rewarded with bitcoins and
optional (voluntary) transaction fees. The additional  rewarded bitcoins are the mechanism that
increases the bitcoin money supply.

For our empirical study, we use transaction-level data with trader identifiers for the Mt.~Gox bitcoin exchange. We conduct our analysis over the uninterrupted period from June 26, 2011 to November 29, 2013. Mt.~Gox was the leading bitcoin trading platform during that period and processed the majority of trading orders.

We extract the data from the Mt.~Gox database leak of March 2014, following Mt.~Gox suspension of its operation and bankruptcy filing. This data set is available on the BitTorrent network and includes a history of all executed trades. The data is organized as a series of comma-separated files with each row listing a time stamp, a trade ID, a user ID, a transaction type (buy or sell), the currency of the fiat leg, the fiat and bitcoin amounts, and the fiat and bitcoin transaction fees. A subset of the trades additionally reveals the country and state of residence of the user. We ignore these last pieces of information as they are only available for a limited number of trades. A heuristic analysis of trade IDs reveals that they correspond to the concatenation of a POSIX timestamp and a microsecond timestamp. We parse the timestamps accordingly to define the execution time of each trade with a microsecond precision.

The respective legs of the trades are split across multiple lines. We initiate the cleaning procedure by aggregating trade entries according to their trade IDs. We filter out trades whose bid leg or ask leg are missing, and remove all duplicates. We also remove from the sample trades for which the same user identifier appears on both legs. Those trades are either due to a bug in the order book matching algorithm, or are simple data errors. Finally, we only consider U.S. dollar-denominated trades and filter out trades whose fiat amount is smaller than \$0.10 to avoid numerical errors in the computation of the price. We define the tick-time price series as the ratio of the bitcoin amount over the fiat amount for the chronological trades series, rounded to the third decimal.

We confirm the authenticity of the remaining data by comparing them to the data set published by Mt.~Gox in 2013 and its subsequent updates. However, the comparison also reveals two problems related to multi-currency trades.\footnote{On August 27, 2011, Mt.~Gox implemented a form of order book aggregation across currencies, with the exchange acting as intermediary. For exemple, a market buy order in USD could match a limit sell order in EUR, triggering a pair of trades between the users and Mt.~Gox. The two legs share the same trade ID, which allows us to identify them easily. The published data set further distinguishes the primary and non-primary legs of a multi-currency trade. The primary leg is the one where Mt.~Gox is selling bitcoins in exchange for fiat. All missing trades are non-primary legs.} First, 92,174 trades have a systematic data error whereby the fiat amount is the same in the primary and the secondary currency, and thus incorrect by a factor corresponding to the exchange rate between the two currencies. We correct this error by copying the fiat amount from the published data set and updating the price. Second, 129,081 trades corresponding to secondary legs of multi-currency trades are missing from the data set, representing less than 2\% of all trades. We find in unreported robustness checks that the impact of the missing trades have a negligible effect on our results.

A visual analysis of the remaining tick data reveals frequent outliers on the whole time period. We eliminate obvious data errors such as trade prices reported at zero or above \$10,000. We fetch daily high and low prices from the external data source Bitcoin Charts\footnote{See http://www.bitcoincharts.com.} and remove trades whose exchange price lies outside of the high-low interval with a 20\% margin. We also discard `bounceback' outliers as defined in \cite{AitsahaliaMyklandZhang11}. The resulting set of trades is used for our analysis of the bitcoin market.

The data set only includes information on executed trades. It lacks limit orders, and consequently provides no explicit information on the bid-ask spread across time or the depth of the order book. The published data set provides an additional field specifying whether orders are initiated by the buyer or the seller, that is, if they are aggressive bids or aggressive asks. This recording is important for our analysis of the potential determinants of jump occurence. We define the best bid series as the price series of aggressive ask orders, and the best ask series as the price series of aggressive bid orders. In the rare occurrences where the best bid price gets higher than the best ask price, we update the best ask to the value of the bid price; reciprocally, we update the best bid price if the best ask price crosses it.

We construct calendar-time price series by computing the median of the tick-time prices within each interval of 5 minutes. In the case where no trade occurs, we propagate the price from the previous period. We build the calendar-time volume series by summing the respective volumes within each interval, and the trades number series by taking the number of trades on each period.

\begin{figure}[tb]
	\centering
	\includegraphics[scale=0.7]{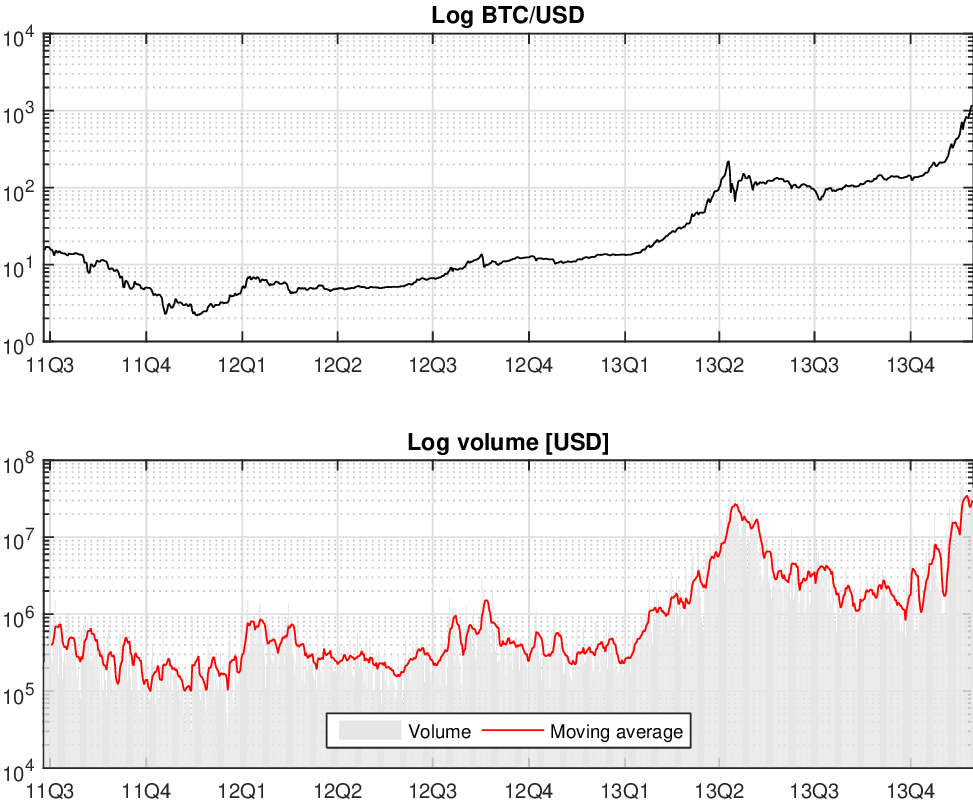}
	\caption[BTC/USD exchange rate and volume]{BTC/USD exchange rate and volume \smallskip \\ \footnotesize{The figures display respectively the bitcoin price in dollar terms and the trading volume at a daily frequency on Mt Gox exchange platform from June 2011 to November 2013.}}
	\label{F:PriceVolume}
\end{figure}

The final data set contains 6.4 million transactions involving 90,382 unique traders. The transactions amount to a total volume of \$2.1 billion, or on average \$2.4 million per day. Figure~\ref{F:PriceVolume} shows the time series of the price and volume on a logarithmic scale during the period. The price of bitcoin increases from \$16 on June 26, 2011 to an all-time high of \$1,207 on November 29, 2013. Volume increases significantly during the period as well, and the linear correlation between price and volume exceeds 70\%. The price of bitcoin has experienced several booms and busts. The clearest example is the crash of April 10, 2013 which saw the bitcoin value drop by 61\% in only hours for no obvious reason, after doubling over the previous week. No stabilizing mechanisms mitigate those large swings. There are no central banks, no market makers, and no circuit breakers in the bitcoin market. 

\section{Methodology}

Many pricing models rely on the assumption that the dynamics of the underlying asset follow a continuous trajectory. For instance, \citet{Black1973} propose a diffusion model with constant volatility and \citet{Heston1993} augments it with a second factor to allow for heteroskedasticity. The empirical literature challenges continuous models \citep[see, e.g.,][]{Ait-Sahalia2002,CarrWu03}. The probability of large moves disappears asymptotically as the horizon shrinks, which does not provide consistent short-term skewness and kurtosis.

There are mainly two approaches to overcome this limitation.\footnote{Another alternative would be to consider Lévy jumps of infinite activity \citep[see, e.g.,][]{Aitsahalia04}.} First, we can introduce a jump component in the price process \citep[e.g.,][]{Merton76,Bates1996}. Jumps are discontinuous price changes occurring instantaneously, no matter the frequency of observations. Alternatively, we can consider models with highly dynamic volatility, such as the two-factor stochastic volatility model of \citet{ChernovGallantGhyselsTauchen03} and \citet{HuangTauchen05}. The probability of sudden moves asymptotically still vanishes, yet those models allow for bursts of volatility leading to significant changes on short-term horizons.

Identifying whether a price process is continuous or has jumps is important because of the implications for financial management such as pricing, hedging and risk assessment. For deep out-of-the-money call options, there may be relatively low probability that the stock price exceeds the strike price prior to expiration if we exclude the possibility of jumps. However, the presence of jumps in the price dynamics significantly increases this probability, and hence, makes the option more valuable. The converse holds for deep in-the-money call options. This phenomenon is exacerbated with short-maturity options. \citet{Barndorff-NielsenShephard06}, \citet{Ait-SahaliaJacod09}, \citet{Mancini01}, \citet{LeeMykland08} develop statistical tools to test for the presence of jumps. Their modeling approach assumes that the data is not contaminated by microstructure noise, preventing a high-frequency analysis. \citet{ChristensenOomenPodolskij14} show that it is crucial to test for jumps at a high frequency to avoid misclassification of bursts of volatility as jumps. \citet{BST15} emphasize the multiple testing issue in jump analysis. After correcting for this bias, they find that jumps are extremely rare events in large-cap stocks.

We follow \citet{LeeMykland12} to test for the presence of jumps in the bitcoin market at a tick frequency. We define a complete probability space $(\Omega,\mathcal{F}_t,\Prob)$, where $\Omega$ is the set of events of the bitcoin market, $\left\{ \mathcal{F}_t : t\in [0,T] \right\}$ is the right-continuous information filtration for market participants, and $\Prob$ is the physical measure. We denote the log-price $P$ and model its dynamics on a given day as
\begin{eqnarray*}
	\der P_t = \sigma \, \der W_t + a Y_t \, \der J_t,
\end{eqnarray*}
where $W_t$ is a Brownian motion, $J_t$ is a jump counting process, $Y_t$ is the size of the jump, $\sigma$ is the volatility assumed to be constant on a one-day period, and $a$ is $0$ under the null hypothesis of no jump and $1$ otherwise.\footnote{We omit the drift term in our log-price model as it has no impact in the jump detection test asymptotically, as explained in Mykland and Zhang (2009).}

The log-price $P$ stands for the unobservable, fundamental price in an ideal market. The bitcoin market is relatively illiquid and is subject to multiple frictions such as trading fees. Consequently, the observed price is contaminated by noise. We define the observed price $\tilde{P}$ as
\begin{eqnarray*}
\label{E:Price}
	\tilde{P}_{t_i} = P_{t_i} + U_{t_i},
\end{eqnarray*}
where $t_i$ is the time of observation\footnote{We assume that Assumption A of \citet{LeeMykland12} about the density of the sampling grid holds.}, $i=1,...,n,$ with $n$ being the number of observations per day. Here $U$ denotes the market microstructure noise with mean $0$ and variance $q^2$. Figure~\ref{F:AutoCorrelation} shows the autocorrelation function at a tick frequency of the observed log-returns on June 10, 2013.\footnote{We observe a similar pattern of significantly negative 1--3 lag coefficients throughout the sample.} The significant dependence in the first lags suggests that the microstructure noise has serial dependence. We therefore allow $U$ to have a $(k-1)$-serial dependence, with $k = 4$.

\begin{figure}[tb]
	\centering
	\includegraphics[scale=0.5]{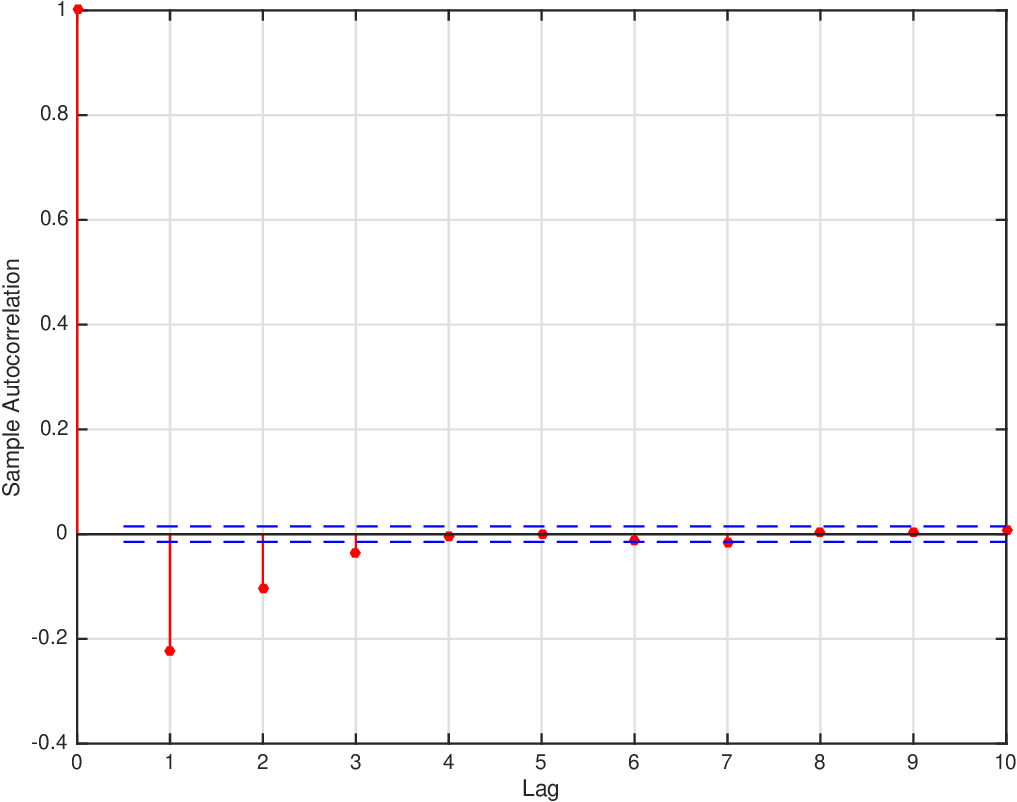}
	\caption[Autocorrelation of log BTC/USD returns]{Autocorrelation of log BTC/USD returns for June 10, 2013 \smallskip \\ \footnotesize{The figure displays the autocorrelogram of the bitcoin price series on June 10, 2013. Dashed horizontal lines show the $5\%$-confidence levels. The autocorrelation is significant up to order $3$.}}
	\label{F:AutoCorrelation}
\end{figure}

We define the block size as $M=\floor{C (n/k)^{1/2}}$, where $\floor{x}$ denotes the integer part of the number $x$, and follow the recommendations of \citet{LeeMykland12}, Section 5.4, for specifying the parameter $C$. We compute the averaged log-price over the block size $M$ as
\begin{eqnarray*}
	\hat{P}_{t_j} = \frac{1}{M} \sum_{i=\floor{j/k}}^{\floor{j/k}+M-1}\tilde{P}_{t_{ik}},
\end{eqnarray*}
and test for the presence of jumps between $t_{j}$ and $t_{j+kM}$ using the asymptotically normal statistic $\mathcal{L}$ defined as
\begin{eqnarray*}
	\mathcal{L}(t_j) = \hat{P}_{t_{j+km}}-\hat{P}_{t_{j}},
\end{eqnarray*}
for $j = 0, kM, 2kM, \ldots$

The asymptotic variance of the test statistic is given by $V = \lim_{n \rightarrow \infty } V_n = \frac{2}{3} 0.2^2 \sigma^2 T + 2 q^2$ where the limit holds in probability. We estimate the volatility $\hat{\sigma}$ using the consistent estimator of \citet{PodolskijVetter2009}, which is robust to the presence of noise and jumps. We use Proposition 1 of \citet{LeeMykland12} to estimate the noise variance $\hat{q}^2$, that is,
\[
	\hat{q}^2 = \frac{1}{2(n-k)}\sum_{m=1}^{n-k}(\tilde{P}_{t_m} - \tilde{P}_{t_{m+k}})^2.
\]
Our estimate of the asymptotic variance is therefore $\hat{V}_n =  \frac{2}{3} 0.2^2 \hat{\sigma}^2 T + 2 \hat{q}^2$.

\citet{LeeMykland12} show the convergence in distribution of the test statistics
\begin{eqnarray*}
	B_n^{-1} \left( \frac{\sqrt{M}}{\sqrt{V_n}} \max_j{|\mathcal{L}(t_j)|} - A_n \right) \longrightarrow \xi,
\end{eqnarray*}
for $j = 0, kM, 2kM, \ldots $, where $\xi$ follows a standard Gumbel distribution with cumulative distribution function
$\Prob (\xi\leq x) = \exp \left( -e^{-x} \right)$, and the constants are as follows
\begin{eqnarray*}
	A_n &= & \left( 2 \log \floor[\bigg]{\frac{n}{kM}} \right)^{1/2} - \frac{\log(\pi) + \log \left( \log \left( \floor[\big]{\frac{n}{kM}} \right) \right)}{2 \left( 2\log \left(\floor[\big]{\frac{n}{kM}}\right) \right)^{1/2}}, \\
	B_n &= & \frac{1}{\left( 2\log \left( \floor[\big]{\frac{n}{kM}} \right) \right)^{1/2}}.
\end{eqnarray*}

We test the presence of jumps on a given day by identifying a divergence of the test statistic from the Gumbel distribution. As emphasized in \citet{BST15}, it is crucial to account for multiple testing when applying a statistical test more than once. Indeed, if the rejection threshold is fixed, the proportion of rejections converges to the size of the test under the null hypothesis because of type I errors, preventing any statistical inference. The FDR ensures that at most a certain expected fraction of the rejected null hypotheses correspond to spurious detections. The FDR approach results in a threshold for the $p$-value that is inherently adaptive to the data. It is higher when there are few true jumps, i.e., the signal is sparse, and lower when there are many jumps, i.e., the signal is dense. Setting the FDR target parameter to $0$ is equivalent to a strict control of the family-wise error rate. It is very conservative as it asymptotically admits no spurious detection due to multiple testing. We prefer a FDR target level of 10\%, which results in a more liberal threshold than with family-wise error rate control. The power of the test is therefore improved, at the cost of accepting that up to 10\% of detected jump days may be spurious. We refer to \citet{BSW10} and \citet{BajgrowiczScaillet12} for further discussion, background, and applications of the FDR methodology in finance (see also \cite{HarveyLiuZhu16} for multiple testing issues in factor modeling).

\section{Empirical results}

In this section, we study the dynamics of jump arrivals on the bitcoin market. We aim to assess the presence of jumps and their distributional properties. We qualify market conditions favoring the apparition of discontinuities and show that jumps have a positive impact on market activity and illiquidity.

\subsection{Jump distribution}
\label{S:JumpDistribution}

We apply the high-frequency jump detection test of \citet{LeeMykland12} with FDR control at a 10\% level and find 124 jump days in the period June 2011 to November 2013, or approximately one jump date per week. Table~\ref{T:Summary} reports the summary statistics for the jumps detected from 5-min intervals and Figure~\ref{F:JumpSizeDistribution} shows the histogram of jump sizes. In 70 cases, the jump has a positive size, and in 54 cases, a negative size. This contrasts with the common idea that jumps depict mainly price crashes. The average size of a positive jump is 4.7\%, and that of a negative jump is $-4.1\%$. We observe discontinuities of up to a 32\% move within a 5-min interval, emphasizing the importance of modeling jumps on this market. Figure~\ref{F:PValues} shows the $p$-values of the jump test statistics, as well as the 1\% confidence threshold and the FDR threshold. We see that a fixed level of 1\% is too permissive and leads to many spurious detections. Interestingly, the thresholding only discards 35\% of rejections, where \citet{BST15} marked up to 95\% as spurious detections on Dow Jones stocks. This is due to the adaptiveness of the FDR control, which is less strict where there are many true jumps in the data.


\begin{table}[tb]
	\centering
	\caption[Summary statistics of jumps]{Summary statistics of jumps \smallskip \\ \footnotesize{The table shows summary statistics for the $124$ jump detections from June 26, 2011 to November 29, 2013. The first column considers all jumps. The second and last columns consider positive and negative jumps, respectively.}}
	\label{T:Summary}
	\begin{tabular}{l c c c} \toprule
		& All jumps & Positive jumps & Negative jumps \\ \midrule
		N & 124 & 70 & 54 \\
		Mean & 0.82\% & 4.65\% & -4.14\% \\
		Mean (abs.) & 4.43\% & 4.65\% & 4.14\% \\
		Med (abs.) & 3.51\% & 3.47\% & 3.52\% \\
		Max & 32.13\% & 32.13\% & -0.76\% \\
		Min & -12.20\% & 1.24\% & -12.20\% \\
		Std dev. & 5.69\% & 4.37\% & 2.43\% \\
		Skewness & 1.33 & 4.05 & -1.09 \\
		Kurtosis & 9.26 & 24.27 & 3.94 \\ \bottomrule
	\end{tabular}
\end{table}

\begin{figure}[t]
	\centering
	\includegraphics[scale=0.5]{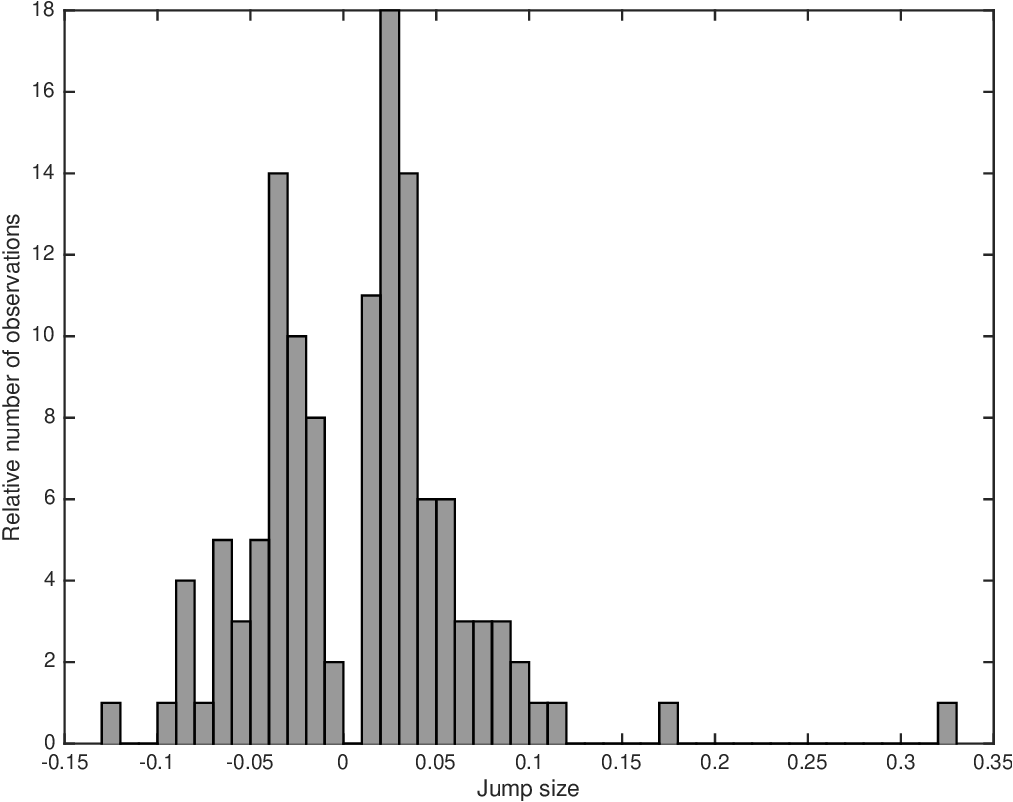}
	\caption[Histogram of the size of jumps]{Histogram of the size of jumps \smallskip \\ \footnotesize{The figure shows the distribution of jump sizes for the $124$ detections from June 26, 2011 to November 29, 2013. In 70 cases, the jump exhibits a positive size. The average size of a positive jump is 4.7\%, and that of a negative jump is 4.1\%. The largest discontinuity is a positive jump of 32\%.}}
	\label{F:JumpSizeDistribution}
\end{figure}

\begin{figure}[t]
	\centering
	\includegraphics[scale=0.5]{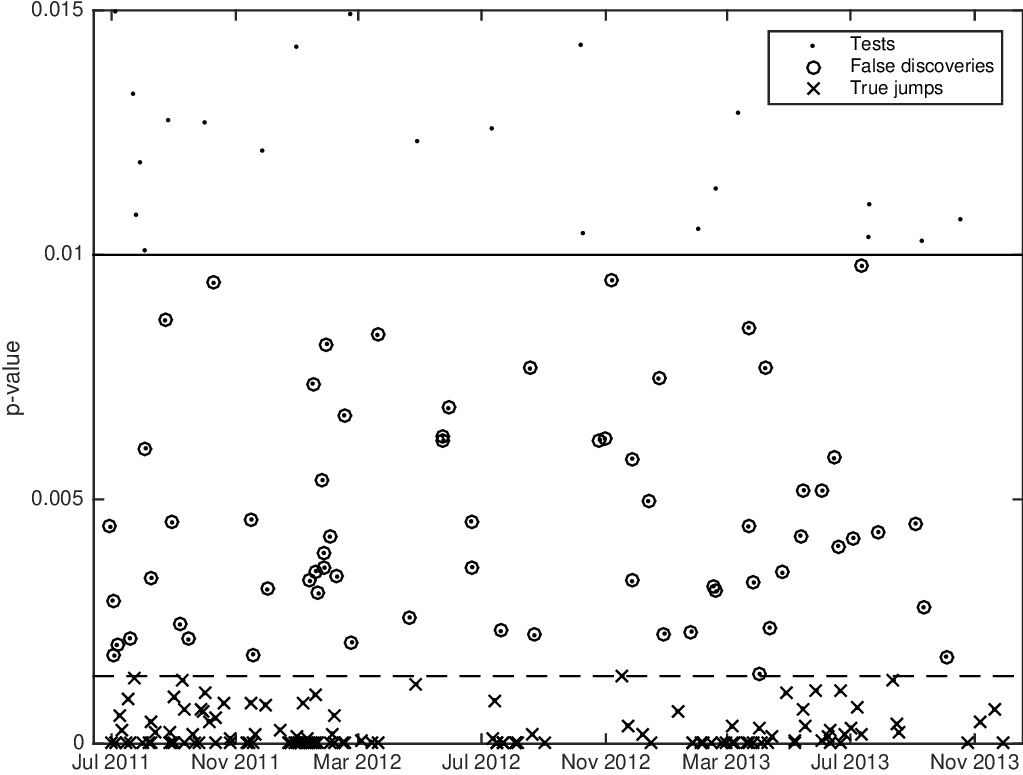}
	\caption[$p$-values of detection statistics]{$p$-values of detection statistics \smallskip \\ \footnotesize{The figure displays the $p$-values of \citet{LeeMykland12} statistics, for every day from June 26, 2011 to November 29, 2013. The solid line indicates the 1\% confidence level and the dashed line indicates the FDR threshold. The 1\%-level is too permissive and leads to many spurious detections due to multiple testing.}}
	\label{F:PValues}
\end{figure}

A widely-used assumption is that jump arrival times follow a simple Poisson process, or equivalently that durations between successive jumps are independent and exponentially distributed. We study the dynamics of jump arrivals to assess whether this assumption is consistent with empirical data. Figure~\ref{F:JumpsPerDay} shows the number of jump detections per quarter on the whole data set. It suggests that the frequency of days with jumps varies across time. Because our test only indicates whether at least one jump occurred on a given date but does not give the exact number of jumps within that day, we cannot test the null hypothesis of exponential inter-jump durations, however. We follow the approach of \citet{BST15} and use the runs test of \citet{Mood40}. The runs test measures the randomness of detections by comparing the number of sequences of consecutive days with jumps and without jump against its sampling distribution under the hypothesis of random arrival. Table~\ref{T:RunsTest} reports the results of the runs test on the full sample and on three sub-periods of 296 days. We strongly reject the hypothesis of independent jump durations on the full sample, indicating significant clustering in jump times. Applying the runs test over three sub-periods reveals that clustering is not equally present on the whole sample. On the period June 26, 2011 to April 16, 2012, which corresponds to the early bitcoin trading days, we observe a strong rejection of the hypothesis of independent runs. On the second period, we only reject at a 10\% level, and we cannot reject on the last period.

\begin{figure}[t]
	\centering
	\includegraphics[scale=0.5]{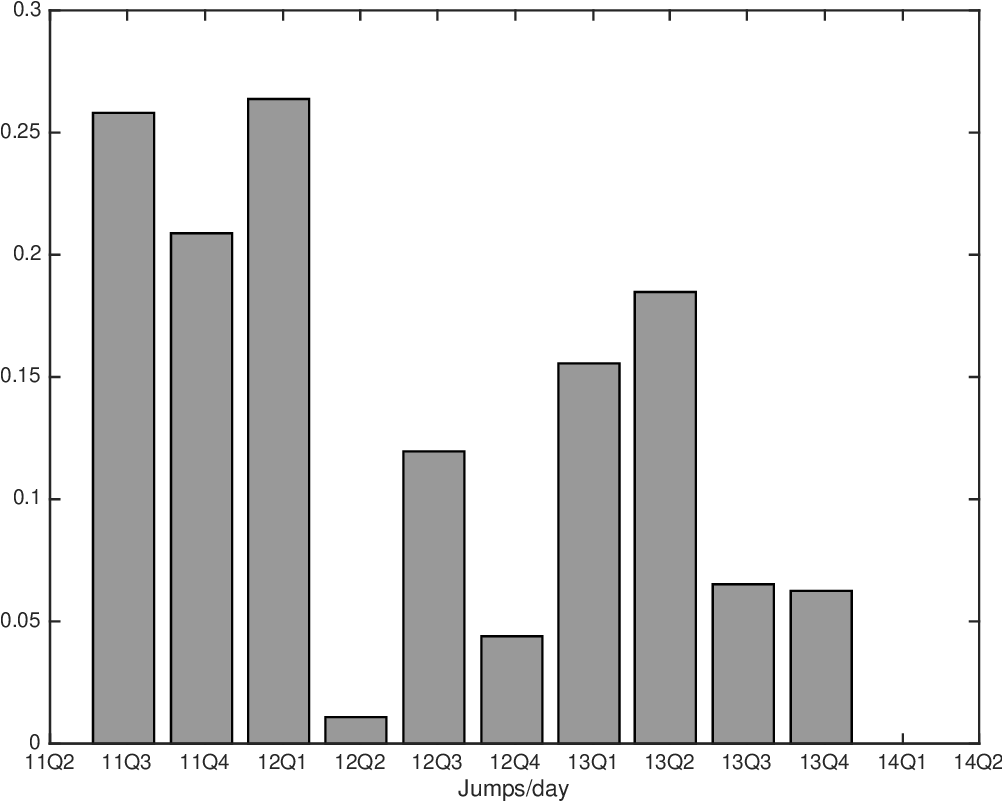}
	\caption[Number of jumps per day across time]{Number of jumps per day across time \smallskip \\ \footnotesize{The figure displays the number of jump detections across time, grouped by quarters.}}
	\label{F:JumpsPerDay}
\end{figure}


\begin{table}[tb]
	\centering
	\caption[Runs test]{Runs test \smallskip \\ \footnotesize{The table shows the results of runs tests applied to jump detections from June 26, 2011 to November 29, 2013, as well as on three sub-periods of equal length.}}
	\label{T:RunsTest}
	\begin{tabular}{l c c c} \toprule
		Period & $p$-value & Jumps & Days \\ \midrule
		Jun 26, 2011 -- Apr 16, 2012 & 0.01 & 67 & 296 \\
		Apr 17, 2012 -- Feb 6, 2013 & 0.09 & 21 & 296 \\
		Feb 7, 2013 -- Nov 29, 2013 & 0.95 & 36 & 296 \\ \midrule
		Entire sample & <0.01 & 124 & 888 \\ \bottomrule
	\end{tabular}
\end{table}

The dynamics of jumps on the bitcoin market contrast with previous literature on high-frequency jump analysis. \citet{BST15} and \citet{ChristensenOomenPodolskij14} identify a small number of jumps on large markets such as Dow Jones constituents, market-wide U.S.\ equity indices and foreign currencies. \citet{BST15} do not identify clustering in the few remaining jumps. We investigate the hypothesis that the relative illiquidity of the bitcoin market coupled with abnormal market activity is key to understanding sudden moves.

\subsection{Jump predictability}
\label{S:JumpPredictability}

Figure~\ref{F:Jump} shows an example of a 5\% positive jump that occurred on June 10, 2013. The highlighted region emphasizes the time interval with the maximum absolute value of $\mathcal{L}(t_j)$ during that day. As illustrated in Panel~(c), the jump occurs after an apparent increase in the trading volume and the order flow imbalance. Panel~(d) also reveals multiple spikes in the bid-ask spread as well as a general widening of the spread shortly before the discontinuity. In this section, we investigate the conjecture that the relative illiquidity of the bitcoin market coupled with abnormal market activity is key to understanding sudden moves. Specifically, we hypothesize that jumps are the result of liquidity drying up in certain market conditions, in conjunction with a regime change in the order flow.\footnote{For a study on the importance of the order flow on price discovery, see, e.g., \citet{Evans2002c}, \citet{Evans2002a}, \citet{Green2004} and \citet{Brandt2004}.}

\begin{figure}[t]
	\centering
	\subfloat[Price\label{F:Jump_Price}]
	{\includegraphics[width=.4\columnwidth]{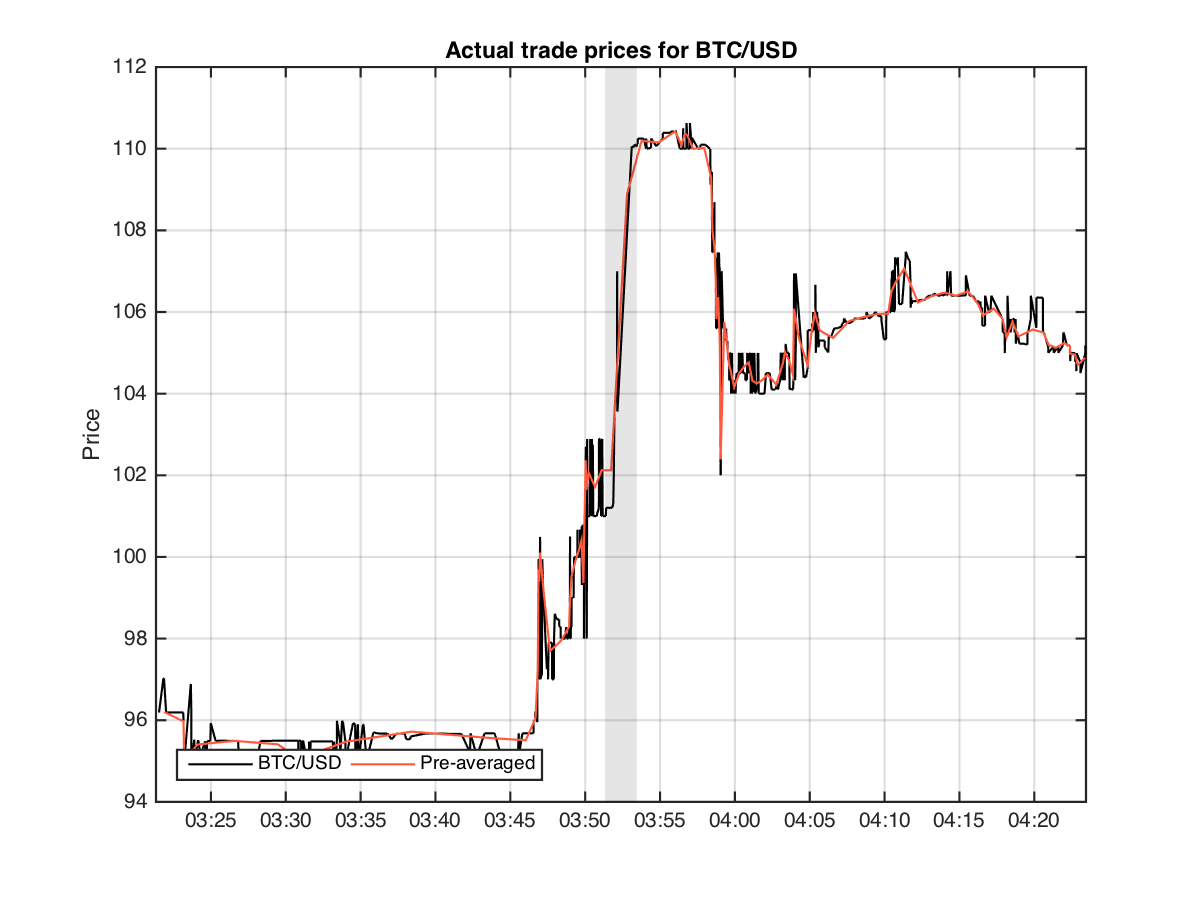}} \quad
	\subfloat[Bids and asks\label{F:Jump_BidsAsks}]
	{\includegraphics[width=.4\columnwidth]{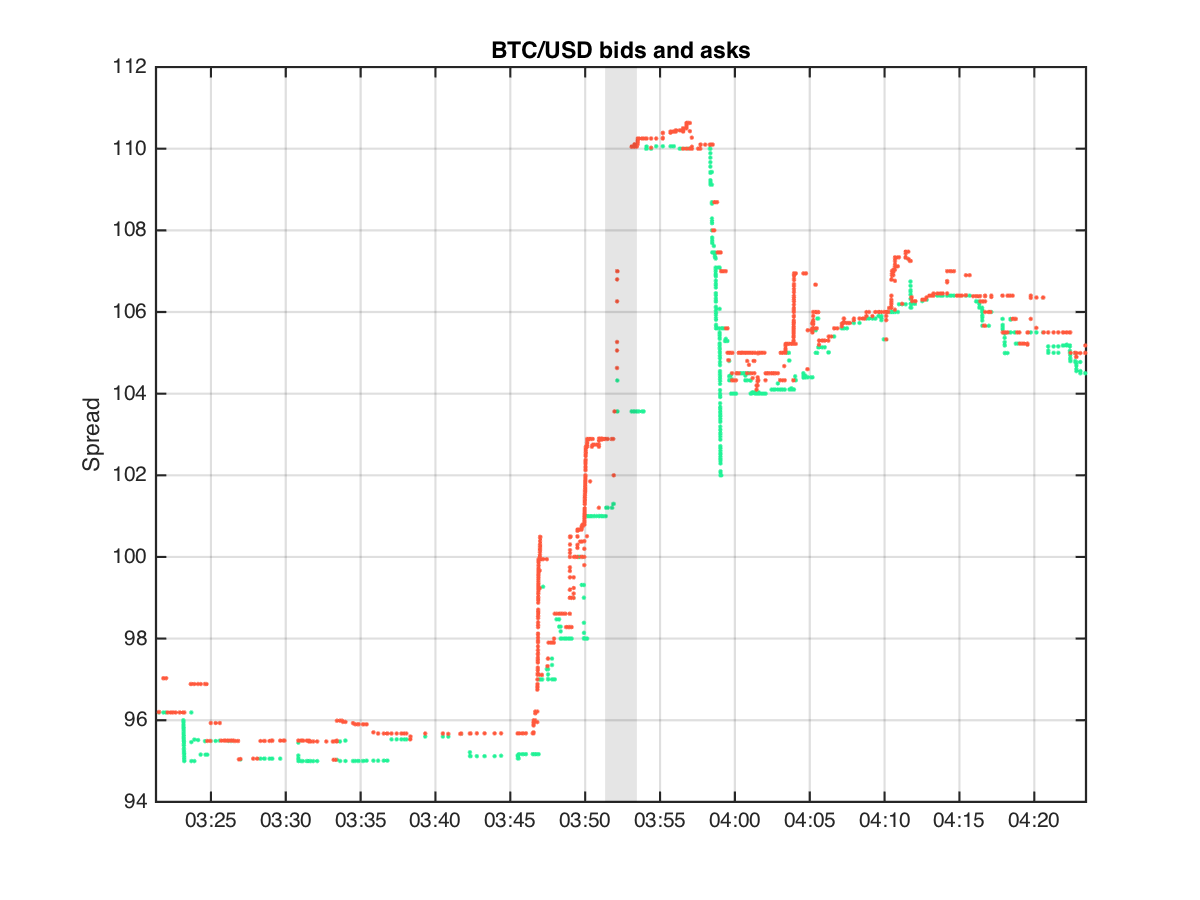}} \\
	\subfloat[Volume\label{F:Jump_Volume}]
	{\includegraphics[width=.4\columnwidth]{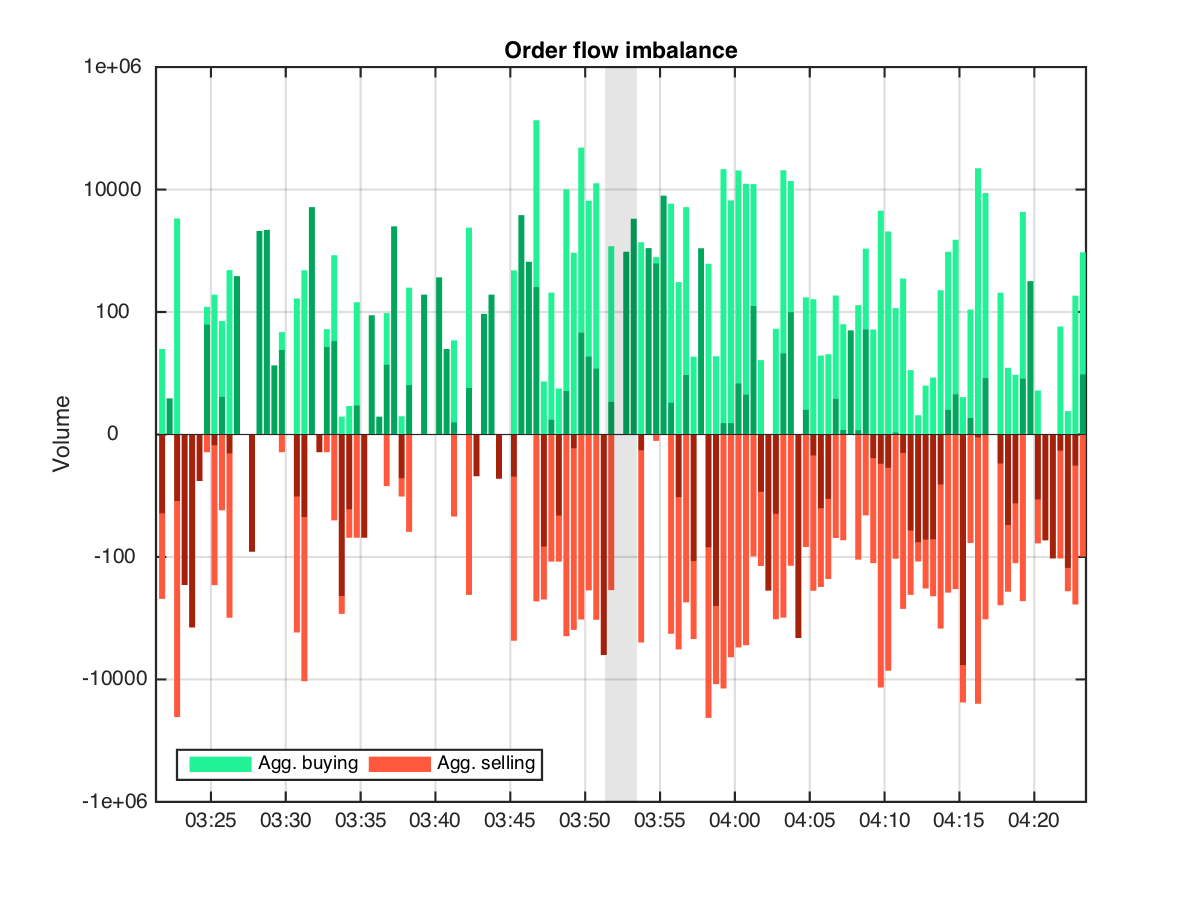}} \quad
	\subfloat[Spread\label{F:Jump_Spread}]
	{\includegraphics[width=.4\columnwidth]{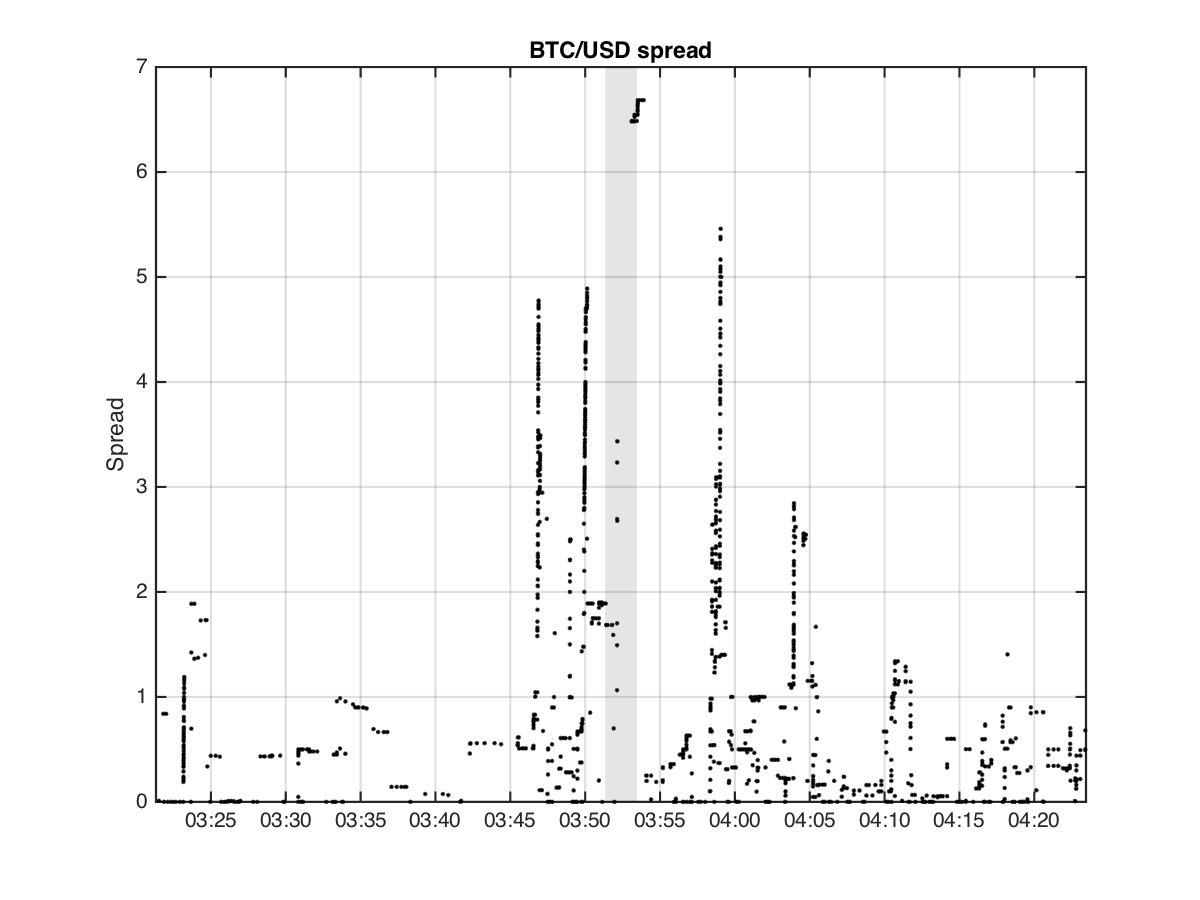}}
	\caption[Jump event of June 10, 2013]{Jump event of June 10, 2013 \smallskip \\ \footnotesize{The panel illustrates the jump detection of June 10, 2013. Panel (a) displays the price series (solid) and the pre-averaged price (dashed). The dark region shows the jump detection period. Panel (b) emphasizes the bid and the ask prices across time. Panel (c) shows the directional volume. Positive (negative) bars count the cumulative volume initiated by aggressive buyers (sellers). Dark bars represent the order flow. Panel (d) shows the evolution of the bid-ask spread.}}
	\label{F:Jump}
\end{figure}

We consider a regular time series at a 5-minute frequency on the whole sample. For each 5-minute period $i$, we set $Y_{t_i} = 1$ if a jump was identified during the period $i$ and $0$ otherwise, and compute the following statistics using the tick data:\footnote{Our results are robust to the choice of frequency. We get similar estimates at a 10-minute and 20-minute frequency, but the significance of estimates decreases strongly at 20-minute. We also try alternative measures of the spread such as the maximum spread on the period or the average spread with no qualitative change. We note that because the jump test of \citet{LeeMykland12} only reveals the largest jump of the day, we might have several time indices $i$ for which $Y_{t_i}$ is incorrectly set to $0$ in the regression.}
\begin{itemize}
	\item $MS_i$ is the bid-ask spread, calculated as the median of the ratio of the bid-ask difference to the mid-price. We use the bid-ask spread factor as a proxy for market illiquidity.
	\item $OF_i$ is the absolute order flow imbalance, defined as the absolute value of the difference between the aggressive buy volume and the aggressive sell volume. A large $OF_i$ thus indicates excessive buying pressure in the market.
	\item $WR_i$ is the `whale'\footnote{The term `whale' is frequently used to describe the big money bitcoin players that show their hand in the bitcoin market. The large players being referred to are institutions such as hedge funds and bitcoin investment funds.} index calculated as the ratio of the number of unique passive traders to the total number of unique traders during the period. The ratio is large when few aggressive traders are responsible for most of the transactions.
	\item $P_i$ is the median observed price.
	\item $RV_i$ is the realized variance of the latent price during the period, given by the noise-robust estimator of \citet{PodolskijVetter2009}.
	\item $NV_i$ is the variance of the microstructure noise, estimated as in \citet{LeeMykland12}.
\end{itemize}
The order flow imbalance $OF_i$ and the whale ratio $WR_i$ quantify two different aspects of the trading pressure that were not directly observable by market participants. The former measures excess directional volume, irrespective of the number of traders responsible for the divergence. For the latter, we take advantage of the richness of our data set that allows us to track the activity of each individually identified trader. The whale index thus gives us a measure of the imbalance between liquidity providers and liquidity takers: a large estimate indicates that few traders are responsible for most of the liquidity taking.


\begin{sidewaystable}[ph!]
	\centering
	\caption[Jump predictability]{Jump predictability \smallskip \\ \footnotesize{The table displays estimates of the probit regression model in Equation \ref{E:Probit}. On Panel A, we compute statistics for periods of 5 minutes. On Panel B, we compute statistics for periods of 10 minutes. First four columns show estimates for the model including fixed effects; last four columns do not include fixed effects. The `Marg. prob.' columns shows the marginal probability change induced by a one-standard deviation change in the values of the covariates from their respective sample averages.}}
	\label{T:Probit}
	\begin{tabular}{l c c c c c c c c} \toprule
		& \multicolumn{4}{c}{With fixed effects} & \multicolumn{4}{c}{Without fixed effects} \\
		\cmidrule(lr){2-5} \cmidrule(lr){6-9}
		Coefficient & Est. & Std error & $p$-value & Marg.\ prob. & Est. & Std error & $p$-value & Marg.\ prob. \\ \midrule

		\multicolumn{9}{c}{Panel A: 5-minute periods} \\ \midrule
		Intercept & -3.61 & 0.13 & <0.01 & & -3.76 & 0.12 & <0.01 & \\
		Realized variance & 2.28 & 6.54 & 0.73 & 4.12\% & 2.10 & 6.37 & 0.74 & 4.50\% \\
		Noise variance & -2876.16 & 876.80 & <0.01 & -56.65\% & -2793.26 & 851.57 & <0.01 & -57.90\% \\
		Abs.\ order flow & 0.00 & 0.00 & <0.01 & 12.16\% & 0.00 & 0.00 & <0.01 & 12.48\% \\
		Med.\ spread & 23.54 & 2.84 & <0.01 & 52.13\% & 23.56 & 2.83 & <0.01 & 52.33\% \\
		Med.\ price & -0.00 & 0.00 & 0.06 & -34.70\% & -0.00 & 0.00 & 0.02 & -34.51\% \\
		Whales & 0.60 & 0.17 & <0.01 & 50.91\% & 0.68 & 0.17 & <0.01 & 43.56\% \\
		Adj.\ $R^2$ & 0.07 & & & & 0.07 & & & \\ \midrule

		\multicolumn{9}{c}{Panel B: 10-minute periods} \\ \midrule
		Intercept & -3.60 & 0.14 & <0.01 & & -3.78 & 0.13 & <0.01 & \\
		Realized variance & -2.54 & 8.09 & 0.75 & -7.76\% & -2.50 & 7.55 & 0.74 & -7.91\% \\
		Noise variance & -2015.06 & 1655.32 & 0.22 & -37.39\% & -1871.53 & 1419.32 & 0.19 & -39.76\% \\
		Abs.\ order flow & 0.00 & 0.00 & 0.01 & 11.58\% & 0.00 & 0.00 & 0.01 & 12.20\% \\
		Med.\ spread & 23.10 & 3.48 & <0.01 & 45.04\% & 22.77 & 3.45 & <0.01 & 46.05\% \\
		Med.\ price & -0.00 & 0.00 & 0.12 & -24.74\% & -0.00 & 0.00 & 0.06 & -25.22\% \\
		Whales & 0.83 & 0.19 & <0.01 & 69.24\% & 0.96 & 0.19 & <0.01 & 58.08\% \\
		Adj.\ $R^2$ & 0.05 & & & & 0.06 & & & \\ \bottomrule
	\end{tabular}
\end{sidewaystable}

We apply a binary probit model to assess the predictive power of these statistics on the probability a jump in the next period and verify our hypothesis. Formally,
\begin{eqnarray} \label{E:Probit}
	\mathbb{P} \left[ J_{i+1} | MS_i, OF_i, WR_i, P_i, RV_i, NV_i \right]  & = &  \Phi \big( \beta_0 + \beta_11_{297:592,i} + \beta_21_{593:888,i}  + \beta_{MS}MS_i + \beta_{OF}OF_i  \nonumber \\  & & + \beta_{WR}WR_i + \beta_{P}P_i + \beta_{RV}RV_i + \beta_{NV}NV_i \big),  
\end{eqnarray}
where $\Phi$ is the Gaussian cumulative distribution function and $1_{t_1:t_2,i}=1$ if $t_1\leq i\leq t_2$, zero otherwise. We add fixed effects for the same sub-periods as in Section~\ref{S:JumpDistribution} to control for the changing market conditions associated with the rapid development of the market for bitcoin. Table~\ref{T:Probit} exhibits the parameter estimates and their respective significance levels. The adjusted pseudo-$R^2=0.07$ confirms the predictive power of the regression, and the unreported likelihood ratio test rejects the constant model at the $0.1\%$ level.

The estimates for $\beta_{MS}$ and $\beta_{OF}$ are both positive and significant, showing the strong impact of market illiquidity and order flow on jump risk. This confirms the results of \citet{JiangLoVerdelhan11}, who find that illiquidity factors and order flow imbalance play a positive role in the occurrence of jumps in the U.S. Treasury market. The estimate of $\beta_{WR}$ is significantly positive as well, indicating that it is not only an imbalance in volume that increases jump risk, but also an asymmetry in the number of aggressive traders relative to their passive counterparts. For $\beta_P$, it is significantly negative, supporting the intuition that jumps have less probability of occurring as the bitcoin market develops and its size increases. Microstructure noise variance plays a negative role in the occurrence of jumps. 
We can explain the negative sign by the probit model capturing the dominant effect that very large values (or at least above the time series average)  of microstructure noise variance are not being followed by a jump most of the time.  
When the microstructure noise variance is large, the market participants do not get a clear signal of the fundamental value of the asset and do not seem to adjust their expectations in an abrupt way.
Yet, in contrast to \citet{JiangLoVerdelhan11}, realized variance has no significant impact on jump risk. Setting aside the obvious differences between the markets for U.S. Treasuries and bitcoin, we believe that the divergence is explained by our use of robust-to-noise estimators and multiple testing adjustments for jump detection on 5-min intervals. 
The positive impact of the realized variance in their empirical results from jump detection on 5-min intervals for many consecutive days could be a consequence of spurious detections.

Panel~B of Table~\ref{T:Probit} reports the estimation of the same model for periods of 10 minutes. The results are consistent with the estimation with 5-minute periods, albeit less categorical, with a slightly lower adjusted pseudo-$R^2$ and the coefficient for microstructure noise variance losing significance, which again highlights the importance of considering high-frequency data for such an analysis.  Our findings thus indicate that jumps are systematically associated with market conditions characterized by a low level of liquidity and the presence of few large and active directional traders.

\subsection{Jump impact}

We perform a post-jump analysis of the market dynamics. On Figure~\ref{F:Impact}, we plot the average dynamics of the whale index, the bid-ask spread, the noise variance and the absolute order flow around jumps. The graphs show that these measures are affected before and after a jump. The whale ratio surges right before a jump, as shown already in Section \ref{S:JumpPredictability}, but quickly reverts to its previous level. The bid-ask spread and the microstructure noise variance gradually increase and peak right around the jump, followed by a slow reversion. The order flow imbalance massively increases before the occurrence but falls to below-average levels right after that. This figure illustrates the intuition of the previous section about the influence of market forces on price discontinuities: aggressive traders placing massive orders, in conjunction with market illiquidity are a significant signal for the occurrence of jumps.

\begin{figure}[t!]
	\centering
	\subfloat[Whales\label{F:Impact_Whales}]
	{\includegraphics[width=.4\columnwidth]{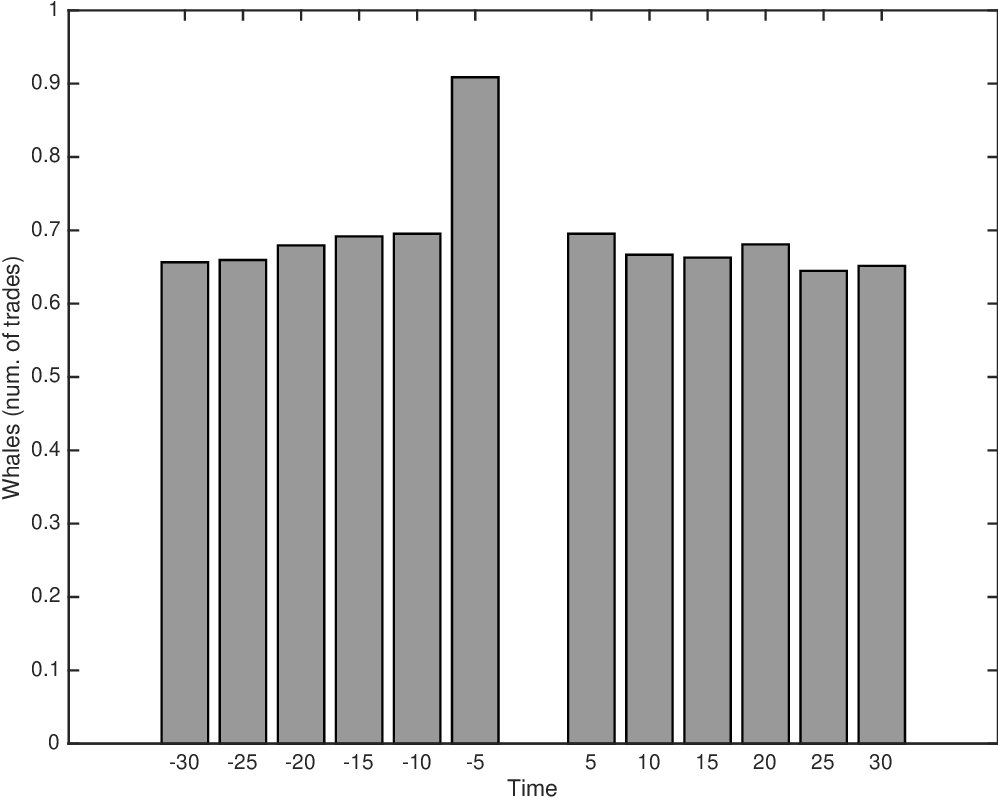}} \qquad
	\subfloat[Spread\label{F:Impact_Spread}]
	{\includegraphics[width=.4\columnwidth]{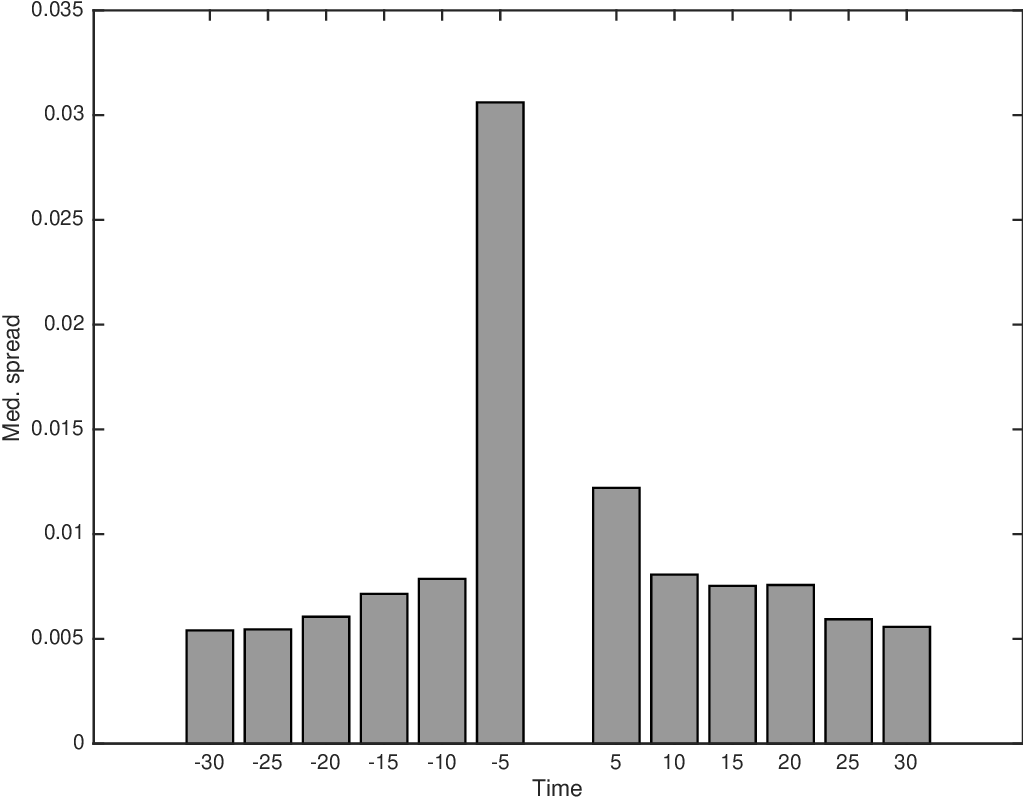}} \\
	\subfloat[Noise variance\label{F:Impact_NV}]
	{\includegraphics[width=.4\columnwidth]{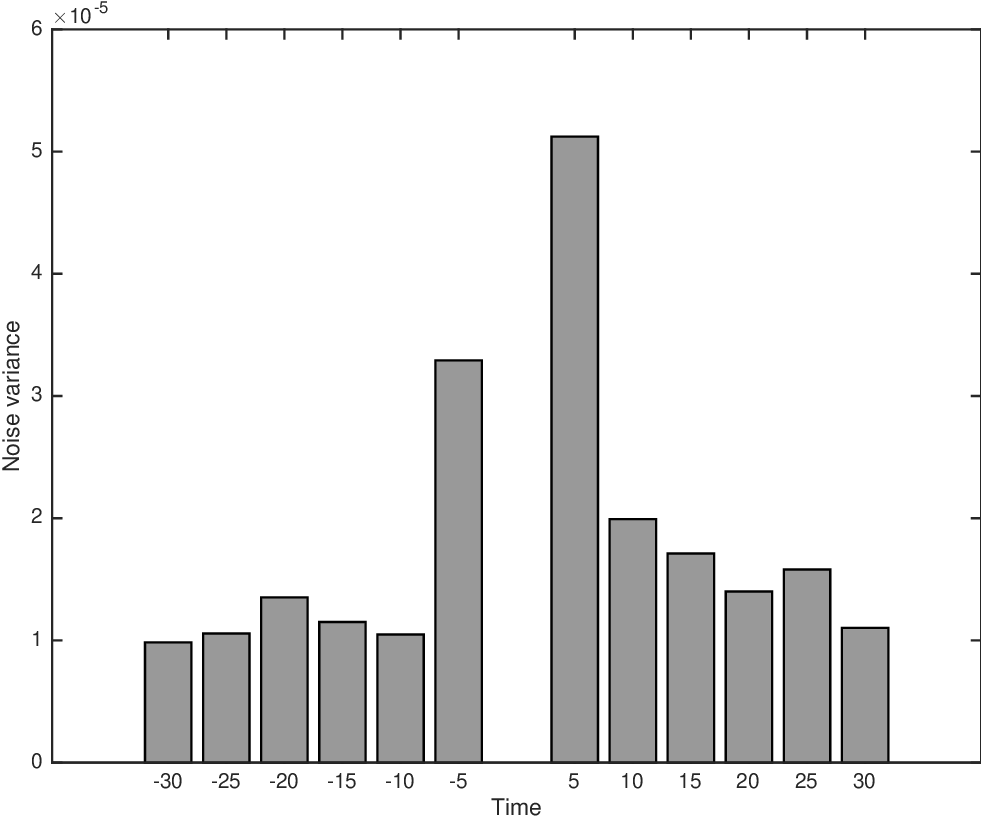}} \qquad
	\subfloat[Order flow imbalance\label{F:Impact_AOF}]
	{\includegraphics[width=.4\columnwidth]{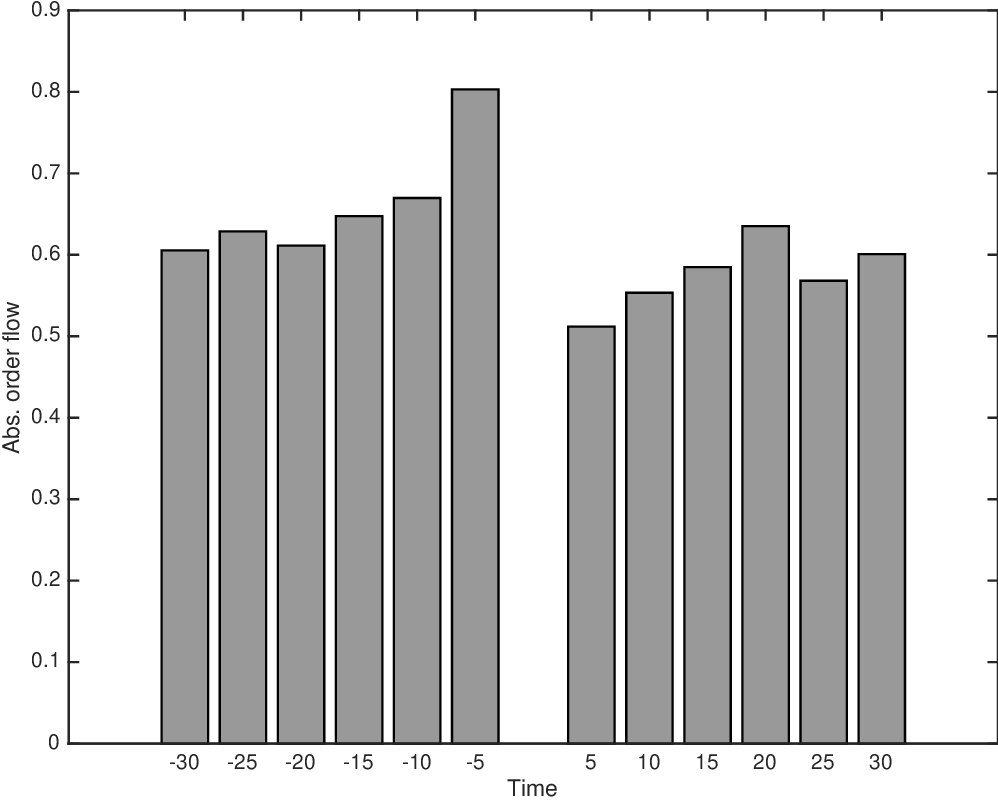}}
	\caption[Market factors around a jump]{Market factors around a jump \smallskip \\ \footnotesize{The panel illustrates four statistics averaged across all jump detections for different periods around jump times. Panel (a) displays the ratio of passive traders over the total number of traders. A high value indicates that few traders are responsible for most liquidity-taking. We observe a clear spike right before the jump occurrence. Panel (b) shows the median bid-ask spread (normalized by the price). We observe a slow widening of the spread punctuated with a large increase before the jump, and a slow reversion to normal levels afterwards. Panel (c) shows the microstructure noise variance, with a significant spike right before and right after a jump detection. The level of the microstructure noise is higher on average after the jump than before. Panel (d) displays the absolute order flow imbalance, which rises sharply before a jump and quickly reverts to normal levels afterwards.}}
	\label{F:Impact}
\end{figure}

The figure emphasizes the market reaction and dynamics after the jumps. We aim to determine if market conditions are affected and how persistent the possible subsequent changes are. We consider the same set of statistics as in the model of Equation~(\ref{E:Probit}), and include additionally the trading volume and the number of traders.  For each jump, we compute the statistics on four consecutive spans of 15 minutes following the detection period. We compare the statistics to a reference period preceding respective jumps by one hour. We define the test statistics as the log-ratio of the post-jump measure over the reference measure for each period. We run a Student $t$-test to assess changes in the means. Table~\ref{T:Impact} gathers the results of $t$-tests, grouped by their respective spans. We find that all measures are exacerbated in the 15 minutes immediately following a jump. The trading volume and the absolute order flow imbalance are abnormally high. At the same time, the number of active traders, and the proportion of aggressive traders are significantly larger. Liquidity proxies including the bid-ask spread and the microstructure noise variance see an increase too, as well as the realized variance. However, the impact of jumps dampens after 30 minutes already. After 45 minutes, all measures revert to anterior levels except the market price: a positive jump generally induces a persistent lower price---and reciprocally, a negative jump induces a higher price. Figure \ref{F:Impact_Price} illustrates this feature by showing the (rescaled) average price around positive and negative jumps, respectively. Jumps tend to occur in episodes of massive price trends and act in an opposite direction to allow for an abrupt and quick price correction. This type of correction is not observed on other markets with stability and liquidity providing mechanisms.


\begin{table}[tp!]
	\centering
	\caption[Jump impact]{Jump impact \smallskip \\ \footnotesize{The table shows the impact of jumps on different market measures. We consider consecutive spans of 15 minutes after jump occurrences and compare a series of statistics for each of them to a reference level preceding the jump by one hour. We apply a Student $t$-test to test the mean of log-ratio statistics for each span. The first two columns include all jump detections, the two central columns include positive jumps only and the last two columns include negative jumps only.}}
	\label{T:Impact}
	\begin{tabular}{l c c c c c c }
		\toprule
		& \multicolumn{2}{c}{All jumps} & \multicolumn{2}{c}{Positive jumps} & \multicolumn{2}{c}{Negative jumps} \\
		\cmidrule(lr){2-3} \cmidrule(lr){4-5} \cmidrule(lr){6-7}
		Statistic & $t$-stat & $p$-value & $t$-stat & $p$-value & $t$-stat & $p$-value \\ \midrule

		\multicolumn{7}{c}{Panel A: 0--15 minutes} \\ \midrule
		Realized variance & 11.22 & <0.01 & 9.27 & <0.01 & 6.54 & <0.01 \\
		Noise variance & 10.04 & <0.01 & 7.60 & <0.01 & 6.56 & <0.01 \\
		Abs.\ order flow & 3.74 & <0.01 & 2.86 & <0.01 & 2.41 & 0.02 \\
		Volume & 6.99 & <0.01 & 5.43 & <0.01 & 4.39 & <0.01 \\
		Num.\ of traders & 7.58 & <0.01 & 6.12 & <0.01 & 4.49 & <0.01 \\
		Med.\ spread & 5.91 & <0.01 & 4.77 & <0.01 & 3.51 & <0.01 \\
		Med.\ price & -2.23 & 0.03 & -3.91 & <0.01 & 1.46 & 0.15 \\
		Whales & 3.10 & <0.01 & 2.94 & <0.01 & 1.37 & 0.18 \\\midrule

		\multicolumn{7}{c}{Panel B: 15--30 minutes} \\ \midrule
		Realized variance & 5.80 & <0.01 & 3.67 & <0.01 & 4.72 & <0.01 \\
		Noise variance & 4.64 & <0.01 & 2.79 & <0.01 & 4.40 & <0.01 \\
		Abs.\ order flow & 2.35 & 0.02 & 1.48 & 0.14 & 2.04 & 0.05 \\
		Volume & 3.60 & <0.01 & 2.62 & 0.01 & 2.48 & 0.02 \\
		Num.\ of traders & 4.01 & <0.01 & 3.40 & <0.01 & 2.19 & 0.03 \\
		Med.\ spread & 2.54 & 0.01 & 1.58 & 0.12 & 2.13 & 0.04 \\
		Med.\ price & -2.04 & 0.04 & -4.27 & <0.01 & 1.71 & 0.09 \\
		Whales & 1.51 & 0.13 & 1.10 & 0.28 & 1.03 & 0.31 \\\midrule

		\multicolumn{7}{c}{Panel C: 30--45 minutes} \\ \midrule
		Realized variance & 3.34 & <0.01 & 2.80 & <0.01 & 1.85 & 0.04 \\
		Noise variance & 2.72 & <0.01 & 2.15 & 0.03 & 1.66 & 0.10 \\
		Abs.\ order flow & 0.58 & 0.56 & 1.28 & 0.21 & -0.57 & 0.57 \\
		Volume & 2.43 & 0.02 & 2.04 & 0.05 & 1.32 & 0.19 \\
		Num.\ of traders & 2.57 & 0.01 & 2.47 & 0.02 & 0.96 & 0.34 \\
		Med.\ spread & 1.45 & 0.15 & 1.62 & 0.11 & 0.17 & 0.86 \\
		Med.\ price & -1.15 & 0.25 & -3.25 & <0.01 & 2.29 & 0.03 \\
		Whales & -0.04 & 0.96 & 0.42 & 0.67 & -0.49 & 0.62 \\\midrule

		\multicolumn{7}{c}{Panel D: 45--60 minutes} \\ \midrule
		Realized variance & 2.04 & 0.02 & 2.07 & 0.03 & 0.68 & 0.28 \\
		Noise variance & 1.85 & 0.07 & 2.11 & 0.04 & 0.28 & 0.78 \\
		Abs.\ order flow & 0.19 & 0.85 & 1.71 & 0.09 & -1.94 & 0.06 \\
		Volume & 2.13 & 0.04 & 2.62 & 0.01 & -0.02 & 0.98 \\
		Num.\ of traders & 1.65 & 0.10 & 2.04 & 0.04 & 0.14 & 0.89 \\
		Med.\ spread & 0.37 & 0.71 & 1.14 & 0.26 & -0.74 & 0.46 \\
		Med.\ price & -0.97 & 0.33 & -2.90 & <0.01 & 2.19 & 0.03 \\
		Whales & -0.62 & 0.53 & 0.36 & 0.72 & -1.38 & 0.17 \\ \bottomrule
	\end{tabular}
\end{table}

\begin{figure}[t]
	\centering
	\subfloat[Negative jumps\label{F:Impact_Price_Neg}]
	{\includegraphics[width=.4\columnwidth]{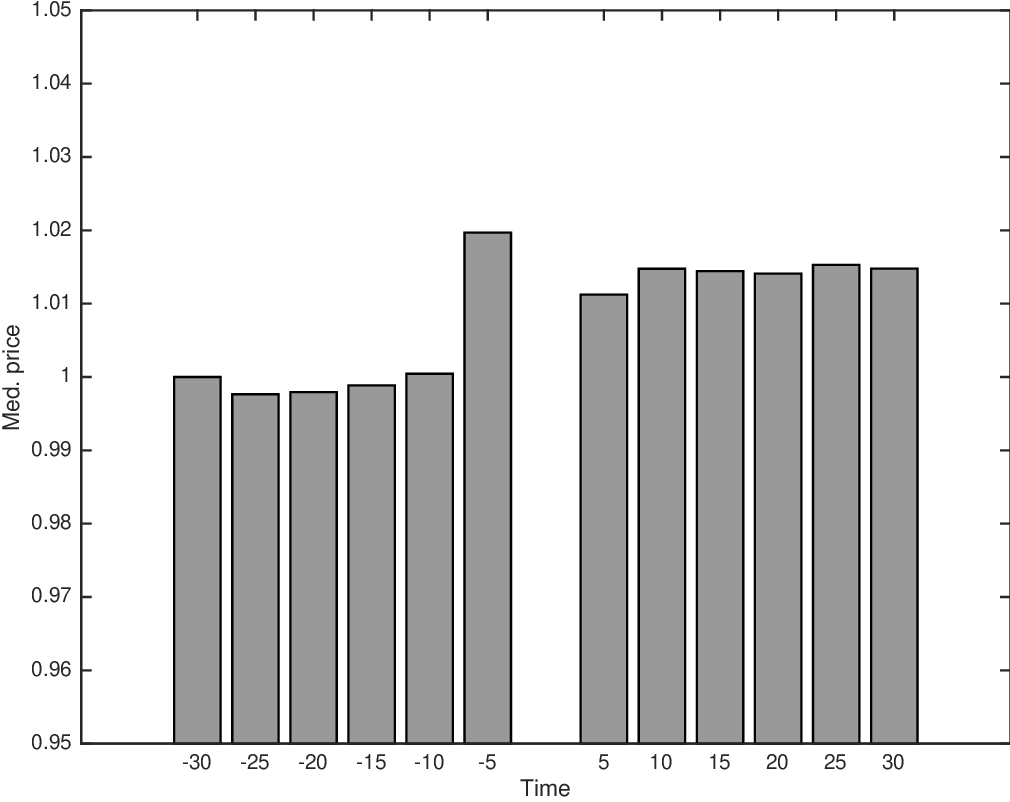}} \qquad
	\subfloat[Positive jumps\label{F:Impact_Price_Pos}]
	{\includegraphics[width=.4\columnwidth]{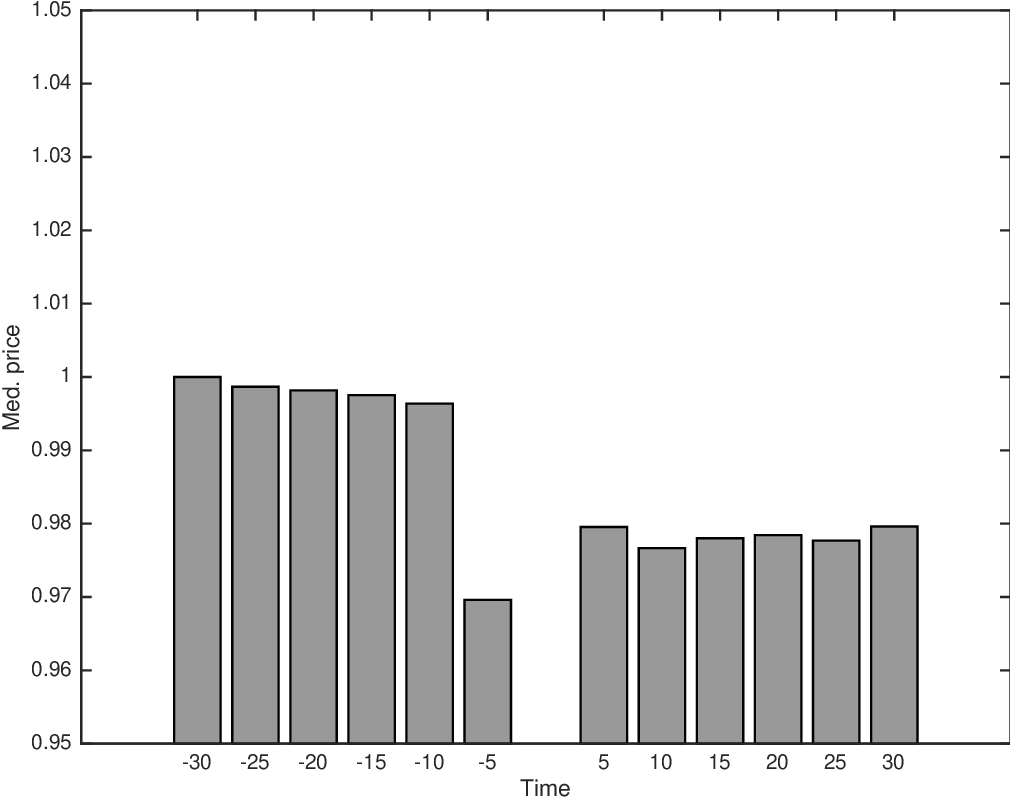}}
	\caption[Jump impact]{Jump impact \smallskip \\ \footnotesize{The panel illustrates the average price levels before and after the occurrence of jumps for a set of 5-minute tranches. Prices are normalized with respect to the price $30$ minutes before the jump to be comparable. Each bar corresponds to the median normalized price on the 5-minute period considered. Panel (a) only considers negative jumps and Panel (b) only considers positive jumps.}}
	\label{F:Impact_Price}
\end{figure}

\section{Conclusion}

The presence of jumps in the dynamics of asset prices remains a debated question in the empirical literature. While many jumps may be detected in low-frequency data, recent studies based instead on high-frequency data have shown that most are in fact misidentified bursts of volatility in continuous price paths. True jumps in large-cap stock prices appear to be rare which prevents systematic studies of their properties.

In this paper, we have been able to conduct such a study for the bitcoin-to-U.S.\ dollar (BTC/USD) exchange rate using transaction-level data obtained from Mt.~Gox exchange , the leading platform during the sample period of June 2011 to November 2013. We contribute to the literature in several ways. First, in contrast to  large-cap stock markets, we find that jumps are frequent: out of the 888 sample days, we identify 124 jump days, or on average one jump day per week. In contrast to the intuition that relates jumps to crash events, most jumps are in fact positive. They are economically significant, with a mean size of 4.65\% for positive jumps and -4.14\% for negative ones. Second, we show that jumps cluster in time: we find runs of jump days that are incompatible with the classical assumption of independent Poisson arrival times. Third, we estimate a binary probit model of jump occurrence using covariates that proxy for illiquidity and market activity, including the `whale' index, a novel measure of the concentration of order flow across traders that exploits a unique feature of our data set which allows us to identify individual traders. We find that illiquidity, order flow imbalance, and the preponderance of aggressive traders are significant factors driving the occurrence of jumps. Finally, we test for the effect of jumps on several market measures and find that jumps have a positive impact on market activity as proxied by volume and number of traders and a negative impact on liquidity. The measured impacts disappear gradually and are no longer significant after an hour, except for the effect on the price level which is persistent.

We have thus shown that jumps are an essential component of the price dynamics of the BTC/USD exchange rate. They are associated with several identified factors, some of which are directly observable from available market data. These conclusions have immediate implications for the modeling of the exchange rate. Further research could seek to verify whether we can extend our conclusions to other financial markets that share characteristics with the studied market, but whose detailed transaction level records are still unavailable.
\newpage
\bibliographystyle{jf}
\bibliography{bibliography}
}

\end{document}